\documentclass[aps,prl,twocolumn,tightenlines,showpacs,aps,amsmath,amssymb,nofootinbib]{revtex4-1}

\usepackage{latexsym}
\usepackage{graphicx}
\usepackage{subfigure}
\usepackage{cases}
\usepackage{hyperref}
\usepackage{amssymb}
\usepackage{bm}
\usepackage{natbib}
\usepackage{amsmath}
\usepackage{slashed}
\usepackage{braket}
\usepackage{color}

\newcommand{\be}{\begin{equation}}
\newcommand{\ee}{\end{equation}}
\newcommand{\ba}{\begin{eqnarray}}
\newcommand{\ea}{\end{eqnarray}}
\def\bal{\begin{align}}
\def\eal{\end{align}}
\def\bald{\begin{aligned}}
\def\eald{\end{aligned}}

\newcommand{\per}{\, .}
\newcommand{\com}{\, ,}
\newcommand{\eref}[1]{Eq.~(\ref{#1})}

\def\L{{\scriptscriptstyle L}}
\def\R{{\scriptscriptstyle R}}
\def\yql{y_{\scriptscriptstyle{Q}}}
\def\yll{y_{\scriptscriptstyle{L}}}
\def\yur{y_{u_{\scriptscriptstyle R}}}
\def\ydr{y_{d_{\scriptscriptstyle R}}}
\def\yer{y_{e_{\scriptscriptstyle R}}}
\def\alphay{\alpha_{\scriptscriptstyle \rm y}}
\def\alphaw{\alpha_{\scriptscriptstyle \rm w}}
\def\alphas{\alpha_{\scriptscriptstyle \rm s}}
\def\Nc{N_{\rm c}}
\def\Nw{N_{\rm w}}

\def\sc{\scriptscriptstyle}
\def\muql{\mu_{{\scriptscriptstyle Q}}}
\def\mull{\mu_{{\scriptscriptstyle L}}}
\def\muur{\mu_{{\scriptscriptstyle u}_{\scriptscriptstyle R}}}
\def\mudr{\mu_{{\scriptscriptstyle d}_{\scriptscriptstyle R}}}
\def\muer{\mu_{{\scriptscriptstyle e}_{\scriptscriptstyle R}}}
\def\edotb{{\bm E} \cdot {\bm B}}
\def\gammay{\gamma_{\rm \scriptscriptstyle y}}
\def\gammaws{\gamma_{\rm \scriptscriptstyle w}}
\def\gammass{\gamma_{\rm \scriptscriptstyle s}}
\def\gammaw{\Gamma_{\rm w}}
\def\gammas{\Gamma_{\rm s}}
\def\etaql{\eta_{{\scriptscriptstyle Q}}}
\def\etall{\eta_{{\scriptscriptstyle L}}}
\def\etaur{\eta_{u_{\scriptscriptstyle R}}}
\def\etadr{\eta_{d_{\scriptscriptstyle R}}}
\def\etaer{\eta_{e_{\scriptscriptstyle R}}}
\def\cy{C_{\rm y}}
\def\cw{C_{\rm w}}
\def\cs{C_{\rm s}}
\def\hinf{H_{\rm inf}}

\newcommand{\uy}{\rm{U(1)}_{\scriptscriptstyle Y}}

\begin{document}

\title{Hypermagnetic Fields and Baryon Asymmetry from Pseudoscalar Inflation}

\date{\today}

\author{Mohamed M. Anber}
\email[]{mohamed.anber@epfl.ch}
\author{Eray Sabancilar}
\email[]{eray.sabancilar@epfl.ch}
\affiliation{Institut de Th\'eorie des Ph\'enom\`enes Physiques, EPFL, CH-1015 Lausanne, Switzerland.}

\begin{abstract}
We show that maximally helical hypermagnetic fields produced during pseudoscalar inflation can generate the observed baryon asymmetry of the universe via the $B+L$ anomaly in the Standard Model. We find that most of the parameter space of pseudoscalar inflation that explains the cosmological data leads to baryon overproduction, hence the models of natural inflation are severely constrained. We also point out a connection between the baryon number and topology of the relic magnetic fields. Both the magnitude and sign of magnetic helicity can be detected in future diffuse gamma ray data. This will be a smoking gun evidence for a link between inflation and the baryon asymmetry of the Universe.
   
\end{abstract}
\pacs{
98.80.Cq,  
12.15.-y.  
      }


\maketitle

\emph{Introduction.}---Baryon asymmetry of the Universe (BAU) remains one of the greatest puzzles of cosmology. As was pointed out by Sakharov, any model that explains BAU has to satisfy three conditions: (1) baryon number non-conservation, (2) C and CP violations, and (3) departure from thermal equilibrium \cite{Sakharov:1967dj}. In fact, many of the early Universe problems were resolved within the inflationary paradigm. In this regard, one may wonder if BAU can also find its resolution within inflation. 

Since its proposal as a dynamical solution to the strong CP problem \cite{Peccei:1977hh}, axions provided a playground for rich phenomenology. These pseudo Nambu-Goldstone bosons (pseudoscalars) appear as a result of a broken global symmetry. Besides, axion-like particles are abundant in string theory and are perfect candidates to build UV complete models of inflation, thanks to their radiatively stable properties. This stability is attributed to the existence of a flat direction that is protected by a shift symmetry. The symmetry eventually gets broken by nonperturbative effects, hence, the desired small values of the slow roll parameters arise in a technically natural way \cite{Freese:1990rb}.

Pseudoscalars are naturally coupled to U(1) gauge fields through the dimension-5 operator $(\Phi/f) Y_{\mu\nu} \tilde Y^{\mu\nu}$, which is inevitable from the effective field theory point of view. This term breaks the conformal invariance of the theory and leads to the production of cosmologically relevant fields\footnote{See also Refs.~\cite{Ratra:1991bn,Dolgov:1993vg,Gasperini:1995dh} for various mechanisms for generating primordial magnetic fields.} \cite{Turner:1987bw}. The fields produced via this mechanism are coherent over the horizon scale and have maximal helicity \cite{Anber:2006xt}. Hence, they carry a non-zero Chern-Simons density, which breaks the macroscopic CP invariance of the Universe.

In this letter, we point out that the change in the Chern-Simons density of the hypercharge field, $\dot h$, produced during inflation feeds into the baryon and lepton number anomaly in the Standard Model, which eventually gets converted into the BAU. C invariance is broken since $\dot h$ enters the anomaly equation with opposite signs for different chiralities (see \eref{anomaly eq} and Table~\ref{tab:coefficients}). In addition, $\dot h$ provides the non-equilibrium condition required by Sakharov's criteria and sustains the BAU. We find that natural inflation leads to overproduction of baryons for most of the parameter space compatible with the cosmological data, hence this class of inflation models are severely constrained. In what follows, we use the metric $\eta_{\mu\nu}={\rm diag}(1,-1,-1,-1)$ and set $c=1$, $\hbar =1$, $k_B =1$. 

\emph{Generation of the hypercharge field.}---We consider a pseudoscalar inflaton $\Phi$ coupled to the $\uy$ hypercharge gauge field $A_\mu$ with field strength $Y_{\mu\nu}$:
\ba
\bald
{\cal L}=\frac{1}{2}(\partial_\mu \Phi)^2-V(\Phi)-\frac{1}{4}Y_{\mu\nu}Y^{\mu\nu}-\frac{\alpha}{4f}\Phi Y_{\mu\nu}\tilde Y^{\mu\nu} \com~~~~
\eald
\ea
where $V(\Phi)$ is the inflaton potential, $f$ is the axion constant, and $\alpha$ is the dimensionless coupling constant between the axion and hypercharge field. The equation of motion of the gauge potential $A_{\mu}$ reads\footnote{\label{ first footnote} Here, we use the radiation gauge setting $A^0=0$ and $\boldsymbol\nabla\cdot \boldsymbol A=0$. In this gauge, the electric and magnetic fields are $\boldsymbol E=-\boldsymbol A'/a^2$ and $\boldsymbol B=\boldsymbol \nabla \times \boldsymbol A/a^2$, where $a$ is the scale factor in the Friedman-Robertson-Walker Universe.} 
\begin{eqnarray}\label{main equation of A}
\left(\frac{\partial^2}{\partial \tau^2}-\nabla^2 -\alpha \frac{\Phi'}{f}\boldsymbol \nabla \times\right)\boldsymbol A=0\,,
\end{eqnarray}
where $\tau$ is the conformal time and the prime denotes the derivative with respect to it.The generation of the hypercharge field can be envisaged by promoting the classical field $A$ to an operator $\hat A$, and then decomposing $\hat A$ into annihilation and creation operators $\hat a_{\lambda}^{\boldsymbol k}$ as: $\hat{\boldsymbol A}=\sum_{\lambda=\pm}\int \frac{d^3k}{(2\pi)^{3/2}}\left[ \boldsymbol\epsilon_{\lambda}A_{\lambda} \hat a_{\lambda}^{\boldsymbol k}e^{i\boldsymbol k\cdot \boldsymbol x}+\mbox{h.c.} \right]$, where the circular helicity vectors $\boldsymbol \epsilon_{\pm}$ obey the relations $\boldsymbol k\cdot \boldsymbol \epsilon_{\pm}=0$ and $\boldsymbol k\times \boldsymbol\epsilon_{\pm}=\mp i|\boldsymbol k|\boldsymbol \epsilon_{\pm}$. Substituting $\hat{\boldsymbol A}$ into (\ref{main equation of A}), we find that the mode functions $A_{\pm}$ satisfy the equation $A''_{\pm}+\left(k^2\pm \alpha k \Phi'/f\right)A_{\pm}=0$ \cite{Garretson:1992vt}.

During the inflationary stage  $d\Phi/dt\equiv\dot\Phi_0$ is constant, where $t$ is the cosmic time. The scale factor is given by $a\cong-1/(H_{\mbox{\scriptsize inf}}~\tau)$, where $H_{\mbox{\scriptsize inf}}=a'(\tau)/a^2(\tau)$ is the Hubble parameter during inflation. Hence, the equations of motion of the modes $A_{\pm}$ read
$\frac{d^2 A_{\pm}}{d\tau^2}+\left[k^2\mp 2k\frac{\xi}{\tau} \right]A_{\pm}=0\,$,
where we defined the dimensionless parameter $\xi\equiv\alpha \frac{\dot\Phi_0}{2fH_{\mbox{\scriptsize inf }}}$. For $k\gg |2\xi/\tau|$ the modes are in their vacuum. However, these modes will develop an instability when $k \sim |\xi/\tau|$.  Depending on the sign of $\xi$, either $A_{+}$ or $A_{-}$ modes will be amplified. For late times, $|k\tau|\ll2\xi$, it is found that \cite{Anber:2006xt}
\begin{eqnarray}\label{produced field}
A_{\pm}\cong\frac{1}{\sqrt{2k}}\left(\frac{k}{2\xi~ a(\tau) H_{\mbox{\scriptsize inf}}}\right)^{1/4}e^{\pi \xi -2\sqrt{2\xi k/[a(\tau)H_{\mbox{\scriptsize inf}}] }} \per ~~~
\end{eqnarray}
Therefore, either $A_+$ or $A_-$ will be amplified by a factor $e^{\pi \xi}$, and thus the produced hypercharge field is maximally helical. This feature is an essential ingredient of the baryogenesis in pseudoscalar inflation. At this point, we emphasize that all the modes produced during inflation get diluted except the last mode that exits the horizon right before the end of inflation. This is also the mode that enters the horizon at the onset of reheating and is the source for the BAU. In the above calculations, we neglected the backreaction of the generated hypercharge field on the inflaton. This is a good approximation as long as the energy density  stored in the generated field is less than the inflaton energy density  $\rho_{\mbox{\scriptsize inf}}=3m_p^2H_{\rm inf}^2/(8\pi)$. Using the mode decomposition described above [see also footnote (\ref{ first footnote})] we find that the energy density stored in the hypercharge field is given by $\langle \boldsymbol E^2+\boldsymbol B^2\rangle_{\rm inf}=6!e^{2\pi\xi}H^4_{\mbox{\scriptsize inf}}/(2^{19}\pi^2 \xi^3)$. Thus, we obtain an upper bound on the inflationary Hubble parameter $H_{\mbox{\scriptsize inf}} \lesssim 64\sqrt{\pi/15}\xi^{3/2}e^{-\pi\xi} m_p$.

The hypermagnetic helicity is defined as ${\cal H} = \int d^3x~ {\bm A} \cdot {\bm B}$, which is proportional to the Chern-Simons number. The rate of change of the helicity density is $h' \equiv -2 \lim_{V\to \infty} (1/V)\int d^3x~ \edotb$. As we will show below, the quantity of interest is $\langle {\bm B} \cdot {\bm \nabla} \times {\bm B} \rangle$. During inflation, the mode decomposition described above yields
\begin{eqnarray}
\label{B curl B}
\langle {\bm B} \cdot {\bm \nabla} \times {\bm B} \rangle_{\rm inf}=\frac{1}{a^5} \int \frac{d^3k|\boldsymbol k|^3}{(2\pi)^3} \left(|A_+|^2-|A_-|^2\right)\,,
\end{eqnarray}
where the integral is over the comoving wave-vectors ${\bm k}$. Using \eref{produced field}, cutting off the integral at $k_c\cong 2\xi H_{\mbox{\scriptsize inf }}a(\tau)$ and setting one of the modes $A_{\pm}$ to zero, as only one of the two modes gets amplified, we find\footnote{Here, we choose the appropriate mode that results in a positive $\langle {\bm B} \cdot {\bm \nabla} \times {\bm B} \rangle_{\rm inf}$. This will ensure that we generate baryons rather than antibaryons.} $\langle {\bm B} \cdot {\bm \nabla} \times {\bm B} \rangle_{\rm inf} = \frac{{\cal I} e^{2 \pi \xi}}{\xi^6} \hinf^5$, where ${\cal I} = 6.8 \times 10^{-4}$.

As soon as reheating starts, the Universe becomes filled with a plasma of relativistic particles. The relevant magnetohydrodynamics (MHD) equations are $\partial_t {\bm E} = {\bm \nabla} \times {\bm B} - {\bm J}$, ${\bm J} = \sigma( {\bm E} + {\bm v}\times {\bm B})$, $\partial_t {\bm B} =- {\bm \nabla} \times {\bm E}$, where ${\bm v}$ is the fluid velocity of the plasma and $\sigma \simeq100 T$ \cite{Arnold:2000dr} is its conductivity. Neglecting the $\partial_t {\bm E}$ term, which remains small in the MHD approximation, we obtain ${\bm E} = \frac{1}{\sigma} {\bm \nabla} \times {\bm B} - {\bm v} \times {\bm B}$, $\partial_t {\bm B} = {\bm \nabla \times ({\bm v} \times {\bm B})}+\frac{1}{\sigma} \nabla^2 {\bm B} $. From these equations, we find \cite{Giovannini:1997eg} $\edotb = \frac{1}{\sigma} {\bm B} \cdot {\bm \nabla} \times {\bm B}$, 
where we used ${\bm B}\cdot {\bm v} \times {\bm B} =0$. On the other hand,  in a plasma with a finite conductivity, the evolution of hypermagnetic field as well as helicity are governed by a competition between the dissipation, $(1/\sigma) \nabla^2 {\bm B}$, and advection, ${\bm \nabla \times ({\bm v} \times {\bm B})}$, terms.  To this end, we recall the magnetic Reynolds number (the ratio between the advection and dissipation terms) ${\cal R}_m\equiv v \sigma/k_p$ where $k_p$ is the physical wave-vector of the mode of interest, which is taken to be the last mode that exits the horizon at the end of inflation: $k_p\cong \frac{H_{\mbox{\scriptsize inf}}}{\xi}\frac{T}{T_{\mbox{\scriptsize rh}}}$, where $T_{\mbox{\scriptsize rh}}$ is the reheating temperature.  Assuming instant reheating, we have $T_{\mbox{\scriptsize rh}}\cong 0.25 \sqrt{m_pH_{\mbox{\scriptsize inf}}}$. Hence, we obtain ${\cal R}_m\cong 400 v\xi \sqrt{m_p/H_{\mbox{\scriptsize inf}}}$, which is much bigger than unity for velocities $v\gtrsim \frac{10^{-5}}{\xi} \sqrt{\frac{H_{\mbox{\scriptsize inf}}}{10^{14}{\rm GeV}}}$. In what follows, we assume that the cosmic fluid velocity is large enough that ${\cal R}_m > 1$ will be unavoidable until the electroweak phase transition. Hence, a turbulent flow will be generated, and the hypermagnetic field will no longer diffuse (see, e.g., Ref.~\cite{Durrer:2013pga} for a review). Thus, $\dot h=-2 \langle \edotb \rangle = - (2/\sigma) \langle {\bm B} \cdot {\bm \nabla} \times {\bm B} \rangle$ at a given epoch after inflation redshifts as:
\begin{eqnarray}\label{hdot}
\dot h = - \frac{2{\cal I} e^{2 \pi \xi}}{\sigma \xi^6} \left(\frac{\hinf}{a}\right)^5\,, 
\end{eqnarray}
where we set  $a_{\mbox{\scriptsize inf}}=1$. In fact, it is this change in the helicity density that plays a pivotal role in baryogenesis by sourcing the $B+L$ anomaly, as we show next.

\emph{Chiral Anomaly in the Standard Model.}---The Standard Model fermions
\ba
Q=\left(\begin{array}{c} u_\L \\ d_\L \end{array}\right) \com ~~ L=\left(\begin{array}{c} e_\L \\ \nu_{e_\L} \end{array}\right)\com ~~ u_\R \com ~~ d_\R \com ~~ e_\R
\ea
exhibit chiral anomaly; namely, the baryon and lepton numbers are anomalous in the Standard Model \cite{'tHooft:1976up}. The anomaly equation for a given fermion species $f$ is:
\ba\label{anomaly eq}
\bald
\partial_{\mu} J_{\sc f}^{\mu} &= \cy^f \frac{\alphay}{16 \pi} Y_{\mu\nu} \tilde Y^{\mu\nu} + \cw^f \frac{\alphaw}{8 \pi} W_{\mu \nu}^a \tilde{W}^{a \, \mu \nu}  \\ 
&+\cs^f \frac{\alphas}{8 \pi} G_{\mu \nu}^b \tilde{G}^{b \, \mu \nu} \com
\eald
\ea
where the coefficients $\cy$, $\cw$ and $\cs$ are given in Table~\ref{tab:coefficients}, and $\alphay$, $\alphaw$ and $\alphas$ are the hypercharge, weak and strong fine structure constants, respectively.
\begin{table}[h]
\begin{center}
\begin{tabular}{|c|c|c|c|}
\hline
 & $\cy$ & $\cw$ & $\cs$ \\
\hline
$\begin{array}{c}Q \end{array}$ 
& $\Nc \Nw \yql^2$ 
& $\Nc$ 
& $\Nw$ \\
\hline
$\begin{array}{c}L \end{array}$ 
& $\Nw \yll^2$ 
& $1$ 
& $0$ \\
\hline
$\begin{array}{c}u_\R \end{array}$ 
& $-\Nc \yur^2$ 
& $0$ 
& $-1$ \\
\hline 
$\begin{array}{c} d_\R \end{array}$ 
& $-\Nc \ydr^2$ 
& $0$ 
& $-1$ \\
\hline
$\begin{array}{c} e_\R \end{array}$ 
& $-\yer^2$ 
& $0$ 
& $0$ \\
\hline
\end{tabular}
\end{center}
\vspace{-15pt}
\caption{\label{tab:coefficients}
Coefficients $C^f$ in \eref{anomaly eq}. The multiplicities $\Nc = 3$ and $\Nw = 2$ take into account the color and weak isospin states of a given family of leptons and quarks, and the hypercharges are $\yql = 1/3\com ~\yll = - 1 \com ~\yur = 4/3 \com ~\ydr = - 2/3\com ~\yer = -2$. The charge conjugates $Q^c$, $L^c$, $u_{\R}^c$, $d_{\R}^c$ and $e_{\R}^c$ have the same coefficients, $C^f$, with all the signs flipped. 
}
\vspace{-10pt}
\end{table}

In the vacuum, the rate of these anomalous baryon/lepton number violating processes are suppressed by $e^{-2S_{E}} = e^{-4 \pi /\alpha}$, where $S_{E}$ is the  action of an instanton associated with this process, and $\alpha$ is the fine structure constant of the corresponding gauge field. In particular, the suppression factors are ${\cal O} (e^{-164})$ and ${\cal O} (e^{-55})$ for the weak and strong interactions, respectively. However, in the early Universe, the dominant contribution to the anomalous process is provided by sphalerons, whose rates can well exceed the expansion rate of the universe \cite{Kuzmin:1985mm}.  Therefore, the weak and strong sphalerons become efficient above the electroweak and QCD scales, respectively, and convert various fermions species into one another while preserving the gauge charges and the baryon minus lepton number, $B-L$. The effect of the hypercharge sector $\uy$ is absent  when the vacuum is trivial; namely, when both hyperelectric and hypermagnetic fields are zero, hence there is no change in the Chern-Simons number of this sector (see however Ref.~\cite{Long:2013tha}). However, a change in the Abelian Chern-Simons number, such as the creation of hypermagnetic field with net helicity, contributes to the anomalous processes. This effect is readily seen by integrating Eq. (\ref{anomaly eq}) to obtain $\Delta N_f=-\cy^f \frac{\alphay}{4 \pi}\int d^4 x \boldsymbol E\cdot \boldsymbol B=\cy^f \frac{\alphay}{8 \pi}\Delta {\cal H}$, where $\Delta {\cal H}$ is the change of the total helicity and $\Delta N_f$ is the change in the baryon number. In fact, it is this integrated version of the anomaly equation that carries the physical information since the anomaly is related to the global structure of the theory. The integrated version tells us that $\Delta {\cal H}$ is converted into $\Delta N_f$ or vice versa. If the Universe during reheating were a bad conductor, then $\Delta {\cal H}$ produced during inflation would survive the reheating to source $\Delta N_f$. If on the other hand, the Universe were a perfect conductor, with infinite conductivity, then we would have $\boldsymbol E=0$ and hence $\Delta {\cal H}= \Delta N_f=0$. This can also be seen from Eq. (\ref{hdot}) as we find $\dot h=0$ by sending $\sigma \rightarrow \infty$. However, the Universe has a finite conductivity, and thus, $\dot h \neq 0$. In fact, $\dot h$ is only suppressed by a factor $\frac{k_p}{\sigma}=\frac{1}{100\xi}\sqrt{\frac{H_{\rm inf}}{ D m_p}}$ compared to the poor conductor case. 

\begin{figure}[t]
\begin{center}
\includegraphics[width=70mm]{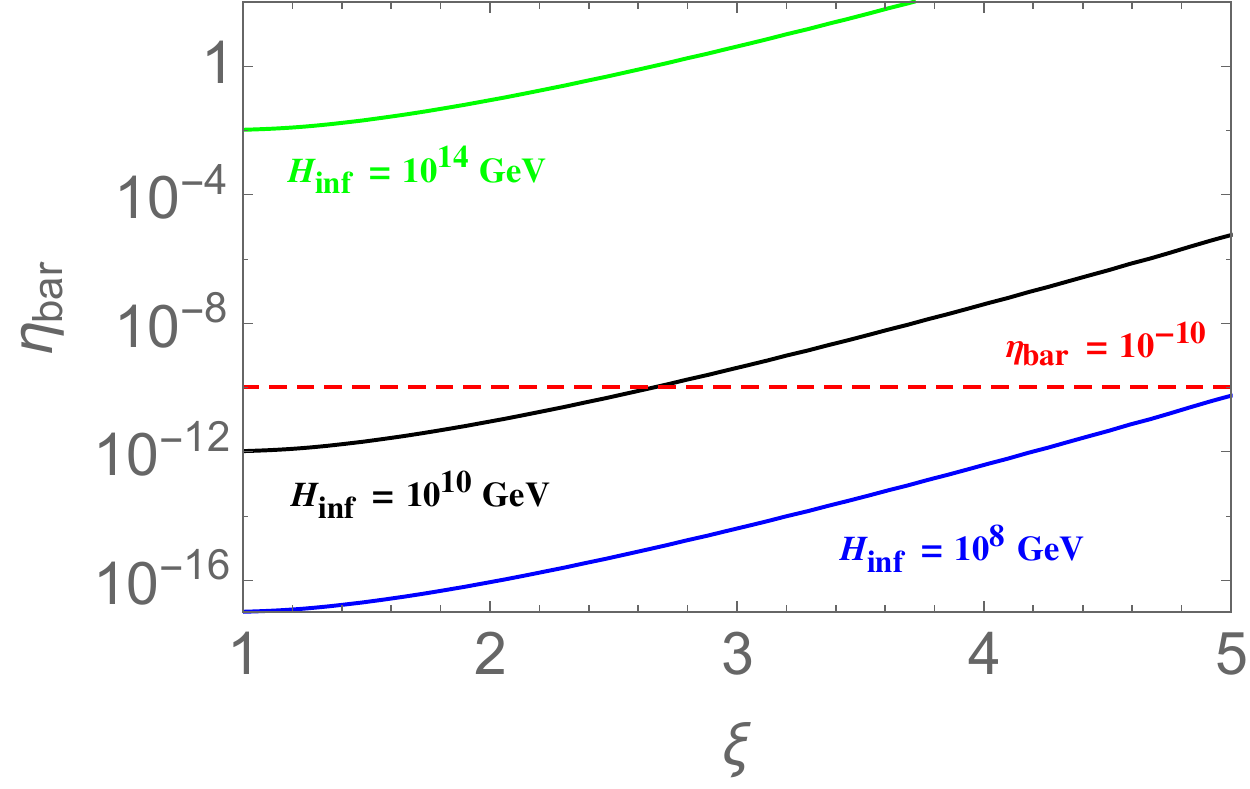}
\includegraphics[width=73mm]{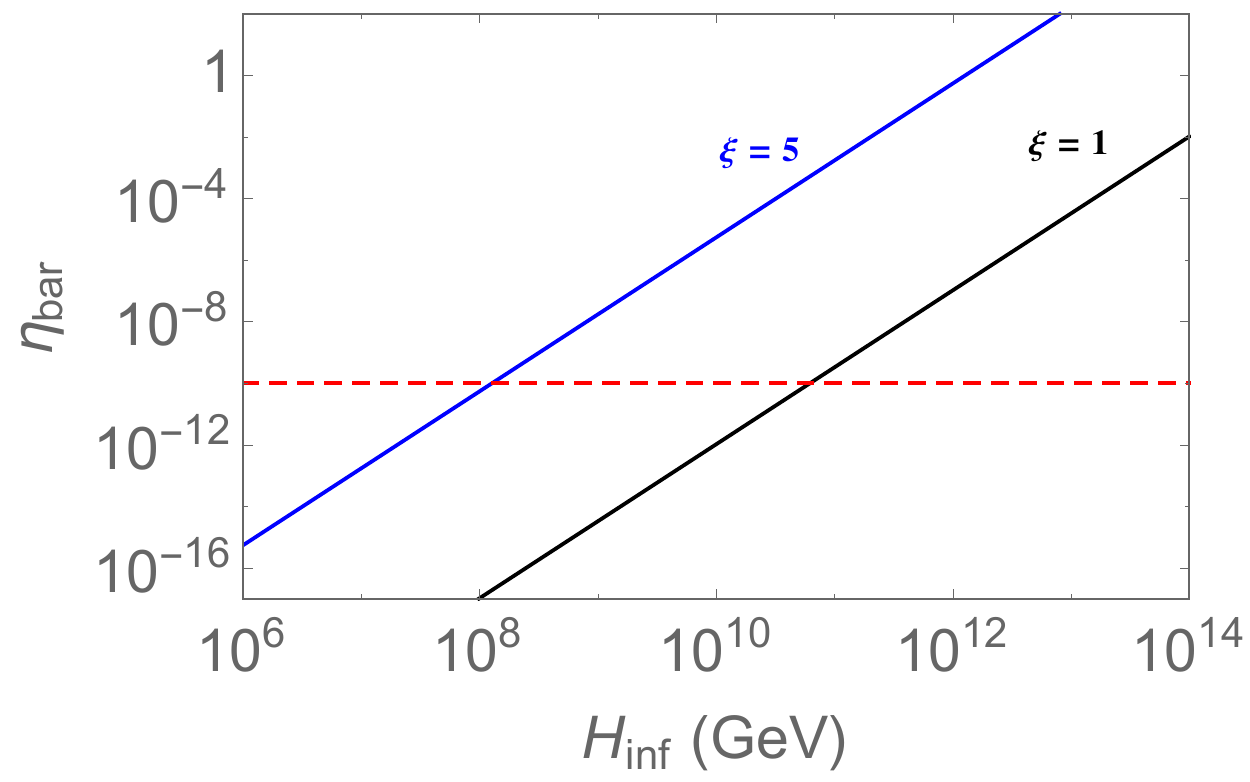}
\vspace{-10pt}
\caption{Top panel $\eta_{\rm bar}$ vs $\xi$, the bottom panel $\eta_{\rm bar}$ vs $H_{\rm inf}$. For instance, the observed value of baryon asymmetry $\eta_{\rm bar} = 10^{-10}$ can be obtained for $\xi=1$ and $H_{\rm inf} =6.3 \times 10^{10}$ GeV.}
\label{fig:eta vs xi}
\end{center}
\end{figure}

\begin{figure}[t]
\begin{center}
\vspace{-20pt}
\includegraphics[width=60mm]{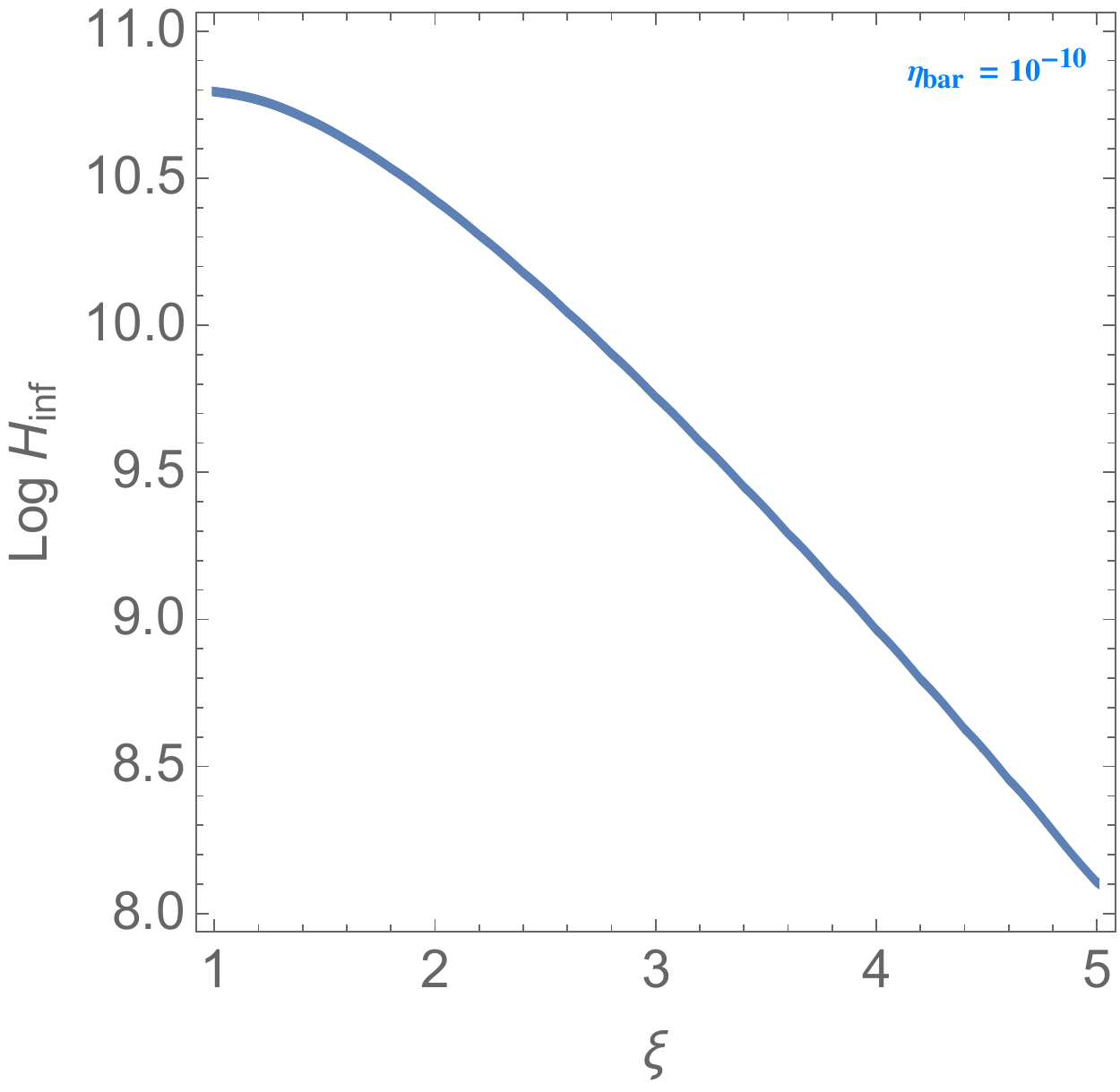}
\vspace{-10pt}
\caption{$H_{\rm inf}$ vs $\xi$ for $\eta_{\rm bar} = 10^{-10}$.}
\vspace{-25pt}
\label{fig:Hinf vs xi}
\end{center}
\end{figure}

     In what follows, we study the effect of the maximally helical hypercharge fields produced during inflation on the evolution of the BAU taking into account the finite conductivity of the Universe during reheating. To this end, we define $\eta_f$ as the asymmetry parameter for a given fermion $f$ as $\eta_f = \frac{n_f - n_{f^c}}{s} \simeq \frac{\mu_f T^2}{6 s}$, where $n_f$ is the number density, $s = 2\pi^2 g_* T^3/45$ is the entropy density, $g_* = 106.75$ is the number of the effective degrees of freedom in the primordial plasma for $T \gtrsim 10^3$ GeV, and $\mu_f$ is the chemical potential. We define the asymmetry parameters for each particle species in the Standard Model as $\etaql \equiv \frac{1}{6s} \Nw \Nc \muql T^2$, $\etall \equiv \frac{1}{6s} \Nw\mull T^2$, $\etaur \equiv \frac{1}{6s} \Nc \muur T^2$, $\etadr \equiv \frac{1}{6s} \Nc \mudr T^2$, $ \etaer \equiv \frac{1}{6s}  \muer T^2 $, and the baryon and lepton asymmetry parameters are $\eta_{\rm bar} = \frac{N_g}{3}( \etaql + \etaur+\etadr)$, $\eta_{\rm lep} = N_g(\etall + \etaer)$, where $N_g=3$ is the number of generations. By taking an ensemble average of the anomaly equation (\ref{anomaly eq}) and taking the difference between the equations of the fields and their charge conjugates, we obtain the Boltzmann equations\footnote{The possibility that baryon asymmetry can be generated from helical primordial hypermagnetic fields via chiral anomaly in the Standard Model was pointed out in Ref.~\cite{Giovannini:1997eg}.} for $\eta_f$  (see for instance Ref.~\cite{Khlebnikov:1988sr} for the derivation of the Boltzmann equation for anomalous charges):
\ba\label{eta_f eqn}
\bald
\frac{\partial \eta_f}{\partial t} &= \cy^f \frac{\alphay}{4 \pi s} \dot h - \cw^f \gammaw (\etaql+\etall) \\
&~~~- \cs^f \gammas (\etaql - \etaur - \etadr) \com ~~~~\\
\eald
\ea
where $t$ is the cosmic time, and we used the relation $\langle Y_{\mu\nu} \tilde Y^{\mu\nu} \rangle = - 4 \langle\edotb \rangle = 2\dot h$. The rates $\gammaw \sim 25 \alphaw^5 T$ \cite{Moore:1997sn} and $\gammas \sim100 \alphas^5 T$ \cite{Moore:1997im} are the weak and strong sphaleron rates, respectively, and the coefficients $C^f$ are given in Table~\ref{tab:coefficients}. In the radiation era we have $H=1/(2t)$, or in terms of temperature $T$, $H = \frac{T^2}{D m_p}$, where $D = \sqrt{\frac{90}{8 \pi^3 g_*}}$. In what follows, it will be more convenient to use the dimensionless variable $x\equiv D m_p /T$ instead of the cosmic time $t$. After making the appropriate change of variables, the Boltzmann equation for the asymmetry parameter $\eta_f$ of a given fermion can be written as a function of $x$:
\ba
\bald
\frac{\partial \eta_f}{\partial x} &= -\cy^f \gammay - \cw^f \gammaws (\etaql+\etall) \\
&~~~- \cs^f \gammass (\etaql - \etaur - \etadr) \com 
\eald
\ea
where we defined $\gammay = \frac{{\cal I} e^{2\pi \xi} \alphay }{100 \xi^6 \sqrt{D}} \left(\frac{\hinf}{m_p}\right)^{5/2}$, $\gammaws = 25 \alphaw^5 \com$ and $\gammass = 100 \alphas^5$. Having obtained the five Boltzmann equations for the five asymmetry parameters, we can readily integrate the system numerically starting from the reheating temperature that corresponds to some inflation scale $H_{\mbox{\scriptsize inf }}$ all the way down to $T\cong 10$ TeV when the weak sphalerons shut off\footnote{In fact, although the $\dot h$ term contributes to the evolution of $\eta_f$ until the electroweak phase transition, $T_{\rm ew} \sim 100$ GeV, for simplicity we assumed that it shuts off at the same temperature as the weak sphalerons. We also neglected the Yukawa terms in \eref{eta_f eqn}, which can only have a minor effect on the final $\eta_f$. }. In Figs.~\ref{fig:eta vs xi} and \ref{fig:Hinf vs xi} we show the results for a range  of the parameter space for $1\lesssim \xi \lesssim5$ taking zero initial values, $\eta_f(T_{\rm rh})=0$. It is important at this point to emphasize that none of the above calculations are robust for $\xi<1$, as in this case one can no longer trust the results for the generated hypercharge field [\eref{produced field}]. The plots in Figs.~\ref{fig:eta vs xi} and \ref{fig:Hinf vs xi} illustrate that the baryon asymmetry is over produced for most of the parameter space. Note also that one should not extrapolate our results for values of $\eta_{\rm bar} \gtrsim 1$ as the perturbative calculation breaks down.
%

\emph{Discussion.}---A simple example of pseudoscalar inflation is natural inflation \cite{Freese:1990rb} in which a shift symmetry $\Phi \rightarrow \Phi+{\cal C}$ is broken down to a discrete subset $\Phi\rightarrow \Phi+2\pi f$ resulting in the potential $V=\Lambda^4\left[1+\cos(\Phi/f) \right]$. We recall that the slow roll conditions are satisfied provided that $m_\Phi^2<H_{\mbox{\scriptsize inf}}^2$. Using Friedmann's equation $3\bar m_p^2 H_{\mbox{\scriptsize inf}}^2=V\cong \Lambda^4$, where $\bar m_p = m_p/\sqrt{8\pi}$, and $m_\Phi^2\cong \Lambda^4/f^2$, we obtain $f>\sqrt{3}\bar m_p$.  The scalar fluctuation power spectrum is given by $P_{\zeta}={\cal P}\left[1+7.5\times 10^{-5}{\cal P}\frac{e^{4\pi \xi}}{\xi^6} \right]
$, where ${\cal P}^{1/2}\equiv \frac{H_{\mbox{\scriptsize inf}}^2}{2\pi |\dot\Phi|}$ \cite{Barnaby:2010vf}. This power spectrum is probed by the CMB observations with amplitude given by the Cosmic Background Explorer (COBE) normalization $P_\zeta\cong 25\times 10^{-10}$. Using the equation of motion of the inflaton during the slow roll regime $3H_{\mbox{\scriptsize inf}}\dot\Phi\cong\partial V/\partial \Phi=\Lambda^4/f$, we find $H_{\mbox{\scriptsize inf}}f\cong 10^{-4}\bar m_p^2$.  Setting $f\sim \bar m_p$, we obtain $H_{\mbox{\scriptsize inf}}\cong 10^{14}$ GeV. Such a large Hubble parameter during inflation will result in baryon asymmetry overproduction for values of $\xi \gtrsim 1$. One way out is that the coupling between the hypercharge gauge field and the inflaton is very weak, $\alpha\ll1$, such that no hypercharge field can be produced during inflation. Such {\em fundamentally} very small values of $\alpha$ appear to be contrived since one expects $\alpha\gtrsim 1$ as a consequence of the "gravity as the weakest force" conjecture \cite{ArkaniHamed:2006dz}. Moreover, a consistent theory of quantum gravity disfavors values of the axion constant $f \gtrsim \bar m_p$. The latter problem can be solved within the framework of N-flation \cite{Dimopoulos:2005ac}. In this scenario, one assumes that there are $N$ different axions with  constants $f_i\equiv f_{\mbox{\scriptsize single}}<\bar m_p$ and that all these axions couple equally to the $\uy$ hypercharge gauge field such that effectively we have $f=\sqrt{N}f_{\mbox{\scriptsize single}}>\bar m_p$ \cite{Anber:2006xt}. Demanding that $f\lesssim \bar m_p$ one finds that $N\sim {\cal O} (100)$ removes the conflict with quantum gravity.  On the other hand, lowering the inflation scale, $H_{\mbox{\scriptsize inf}}\lesssim 6.3 \times 10^{10}$ GeV, guarantees the production of the observed baryon asymmetry for $\xi \gtrsim 1$. Such low inflationary scales will require invoking curvatons in order to respect the COBE normalization \cite{Enqvist:2001zp,Lyth:2001nq,Moroi:2001ct}.

One of the key predictions of natural inflation, apart from the cosmological data, is the maximally helical hypermagnetic field generated via the dimension-5 operator of the form $(\Phi/f) Y_{\mu\nu} \tilde Y^{\mu\nu}$, which is expected from effective field theory considerations. In this letter, we showed that the dramatic consequence of this coupling is the overproduction of baryon asymmetry that severely constrains models of natural inflation with large Hubble parameter, $H_{\inf} \gtrsim 6.3 \times 10^{10}$ GeV. Note, however, that we assumed the primordial plasma to be turbulent between $T_{\rm rh}\gtrsim T \gtrsim10$ TeV. If the plasma ceases to be turbulent during the course of baryon number generation, the contribution from the hypercharge sector might become less efficient due to the diffusion of hypermagnetic field. Meanwhile, weak sphalerons can wash out the excess baryon number. However, we note that this scenario is not likely as the magnetic Reynolds number tends to increase after reheating \cite{Durrer:2013pga}. To conclude, for parameters $H_{\inf} \lesssim 6.3 \times 10^{10}$ GeV and $\xi \sim 1$, the observed BAU can be achieved. Yet another prediction in this case would be the relic magnetic fields with right handed helicity ($ h >0$). It was pointed out in Refs.~\cite{Cornwall:1997ms,Vachaspati:2001nb,Long:2013tha} that the magnetic helicity is proportional to the baryon number. In this work, we explicitly showed this to be case. A recent analysis of the diffuse gamma ray data hints towards a global CP violation, which could be due to primordial magnetic fields with non zero helicity \cite{Tashiro21112014}. It would be a boon to find an observational correlation between the topology of these primordial magnetic fields and the baryon number. This will be a smoking gun evidence for a link between inflation and the BAU.

We would like to thank Kohei Kamada, Andrew J. Long, Oleg Ruchayskiy, Mikhail Shaposhnikov and Tanmay Vachaspati for discussions. We are grateful to Lorenzo Sorbo for enlightening discussions and critical comments on an early version of this manuscript. This work was supported by the Swiss National Science Foundation SNSF.

\bibliographystyle{apsrev4-1}
%


\end{document}